\documentclass[reprint,amsmath,amssymb,aps,prb,]{revtex4-2}

\usepackage{graphicx}
\usepackage{xcolor}

\newcommand{\mybf}[1]{\boldsymbol{\mathbf{#1}}}

\begin{document}
	
\title{Thermal Transport Properties of Magnons on the $\alpha$-T$_3$ Lattice}
	
\author{Luqman Saleem}
\email{luqman.saleem@kaust.edu.sa}
\affiliation{Physical Science and Engineering Division, King Abdullah University of Science and Technology (KAUST), Thuwal 23955, Saudi Arabia
}
\author{Hasan M. Abdullah}
\affiliation{Physical Science and Engineering Division, King Abdullah University of Science and Technology (KAUST), Thuwal 23955, Saudi Arabia
}
\author{Udo Schwingenschlögl}
\email{udo.schwingenschlogl@kaust.edu.sa}
\affiliation{Physical Science and Engineering Division, King Abdullah University of Science and Technology (KAUST), Thuwal 23955, Saudi Arabia
}
\author{Aur\'elien Manchon}
\email{aurelien.manchon@univ-amu.fr}
\affiliation{Aix-Marseille Universit\'e, CNRS, CINaM, Marseille, France
}

\begin{abstract}
We theoretically investigate magnons on the $\alpha$-T$_3$ lattice. Atomistic spin dynamics simulations show that next-nearest neighbor hopping and easy-axis anisotropy stabilize ferromagnetic order in the presence of Dzyaloshinskii-Moriya interaction. We identify one topologically trivial magnon insulator phase and three magnon Chern insulator phases. The topologically trivial magnon insulator phase exhibits a small but non-zero magnon thermal Hall conductivity, while in the magnon Chern insulator phases the Chern number of the lowest magnon band dominates the magnon thermal Hall conductivity. The sign of the magnon thermal Hall conductivity does not change at the topological phase boundaries, but distinct changes are observed in the magnitude. 
\end{abstract}
	
\maketitle
	
\section{Introduction} \label{sec:intro}
Integrating topology with Bloch band theory \cite{Bloch1929} has enabled the discovery of new classes of matter \cite{RevModPhys.88.021004}. Topological insulators \cite{RevModPhys.82.3045, Yan_2012, Tokura2019}, characterized by their robust metallic edge states with insulating bulk, exemplify how topology can lead to exotic electronic properties, such as dissipationless transport \cite{Bauer2016, PhysRevB.105.085412}. Additionally, topology has facilitated the exploration of various quasiparticles, including spinons \cite{Kohno2007,doi:10.1126/science.1201080}, Majorana fermions \cite{PhysRevLett.100.096407, doi:10.1126/science.1259327}, axions \cite{PhysRevLett.102.146805,Nenno2020}, magnetic monopoles \cite{Castelnovo2008,doi:10.1126/science.1167747}, and fractionally charged vortices \cite{PhysRevLett.103.066402}, which hold potential for application in high-performance electronics \cite{RevModPhys.90.015005}, spintronics \cite{Flebus2024,DalDin2024}, and topological quantum computation \cite{PhysRevLett.94.166802,RevModPhys.80.1083,doi:10.1126/science.1231473}.
	
Among the quasiparticles intensively studied by topological band theory are also magnons \cite{Flebus2024}, the quanta of spin waves in crystalline magnetic materials. These bosons without electrical charge are an alternative to electrons for information transport and processing. In particular, high-energy magnons have been intensively studied because they avoid Joule heating \cite{Flebus2024,Kruglyak2010}. Magnons exhibit exotic phenomena such as the thermal Hall effect \cite{PhysRevB.89.134409, PhysRevB.94.174444,PhysRevB.95.014422,PhysRevLett.121.097203,PhysRevB.98.094419,PhysRevB.99.014427}, spin Nernst effect \cite{PhysRevB.93.161106,PhysRevLett.117.217203,PhysRevLett.117.217202,PhysRevB.96.134425,PhysRevB.98.035424,PhysRevB.97.174407,PhysRevB.100.100401,PhysRevResearch.2.013079}, spin Seebeck effect \cite{PhysRevB.81.214418,Uchida2010,PhysRevB.93.014425,PhysRevLett.116.097204}, magnon-mediated Edelstein effect \cite{PhysRevB.101.024427}, and magnon orbital Nernst effect \cite{doi:10.1021/acs.nanolett.4c00430}.
	
Intriguing magnon physics emerges for special crystal lattices. For example, antiferromagnets with Kagome lattice exhibit the spin Seebeck and spin Nernst effects even in the absence of Dzyaloshinskii-Moriya (DM) interaction \cite{Dzyaloshinsky1958,PhysRev.120.91} due to non-collinear alignment of the magnetic moments \cite{PhysRevB.100.100401}. Both ferromagnetic and antiferromagnetic magnons have been studied for the honeycomb \cite{PhysRevB.94.075401,PhysRevX.8.011010,PhysRevX.8.041028,PhysRevX.11.031047,PhysRevB.108.165412, PhysRevB.109.134415,PhysRevB.109.094415,https://doi.org/10.48550/arxiv.2402.14572}, Kagome \cite{PhysRevB.89.134409,PhysRevLett.115.106603,PhysRevB.95.014422,PhysRevLett.121.097203,PhysRevB.98.094419,PhysRevB.99.014427}, Lieb \cite{Cao2015,YARMOHAMMADI2016208,PhysRevB.109.054412}, and triangular \cite{PhysRevB.100.064412} lattices. Theoretical proposals additionally suggest the existence of magnons in the newly discovered class of altermagnets \cite{PhysRevB.108.224421,PhysRevB.108.L180401,PhysRevLett.131.256703,https://doi.org/10.48550/arxiv.2405.05090}.

The $\alpha$-T$_3$ lattice \cite{PhysRevB.34.5208} is interesting due to the similarity of its band structure to that of graphene except for an extra flat band at the Dirac point \cite{PhysRevLett.108.045306,PhysRevA.92.033617,PhysRevB.93.165433}. It is a honeycomb lattice with an additional atom at the center of each hexagon coupled only to one of the other two sublattices with a strength $\alpha>0$. It becomes a honeycomb lattice for $\alpha = 0$ and a dice (or T$_3$) lattice for $\alpha = 1$. Numerous studies have explored the equilibrium and non-equilibrium properties of the $\alpha$-T$_3$ lattice. The $\alpha$-dependent band structure and Berry curvature enable it to host properties such as super Klein tunneling \cite{PhysRevB.84.115136,PhysRevB.93.035422,PhysRevLett.116.245301,PhysRevB.96.024304,PhysRevB.95.235432,PhysRevB.101.165305}, Anderson localization \cite{doi:10.1073/pnas.1620313114,PhysRevB.100.104201}, valley-dependent transport \cite{PhysRevB.96.045418,PhysRevB.99.245412,PhysRevB.103.165114,Niu_2022}, zitterbewegung \cite{Biswas2018}, Ruderman-Kittel-Kasuya-Yosida interaction \cite{PhysRevB.101.235162,PhysRevB.103.075418}, linear and non-linear magneto-optical effects \cite{PhysRevB.92.245410,PhysRevB.99.045420,PhysRevB.100.035440}, Majorana corner states \cite{Mohanta2023}, and Seebeck and Nernst effects \cite{PhysRevB.108.155428}. The topological properties of electrons on the $\alpha$-T$_3$ lattice have been explored in Ref.\ \cite{PhysRevB.109.165118}. The $\alpha$-T$_3$ lattice appears in Hg$_{1-x}$Cd$_x$Te with an intermediate $\alpha=3^{-1/2}$ \cite{Orlita2014,PhysRevB.92.035118} and also has been realized by fabricating a SrTiO$_3$/SrIrO$_3$/SrTiO$_3$ structure \cite{PhysRevB.84.241103}.
	
Despite significant research on the $\alpha$-T$_3$ lattice, the properties of magnons remain unexplored. In this work, we study the $\alpha$-T$_3$ lattice via a Heisenberg spin Hamiltonian with nearest neighbor coupling, next-nearest neighbor coupling, DM interaction, and easy-axis anisotropy. The magnon band structure of the honeycomb lattice is recovered for $\alpha = 0$. We address the magnetic order by atomistic spin dynamics simulations and then extract the topological phase diagram. We calculate the magnon thermal Hall conductivity by determining the Berry curvature. The paper is organized as follows:\ The spin Hamiltonian and its conversion into the magnon Hamiltonian are presented in section \ref{sec:model} followed by the atomistic spin dynamics and thermal transport approaches. The magnetic order along with the magnon band structure, topological phase diagram, and magnon thermal transport properties are discussed in section \ref{sec:results}. Section \ref{sec:summary} gives a summary.
	
\section{Model and method} \label{sec:model}
\subsection{Spin Hamiltonian and Magnon Hamiltonian}
\begin{figure}[b]
\includegraphics[width=\linewidth]{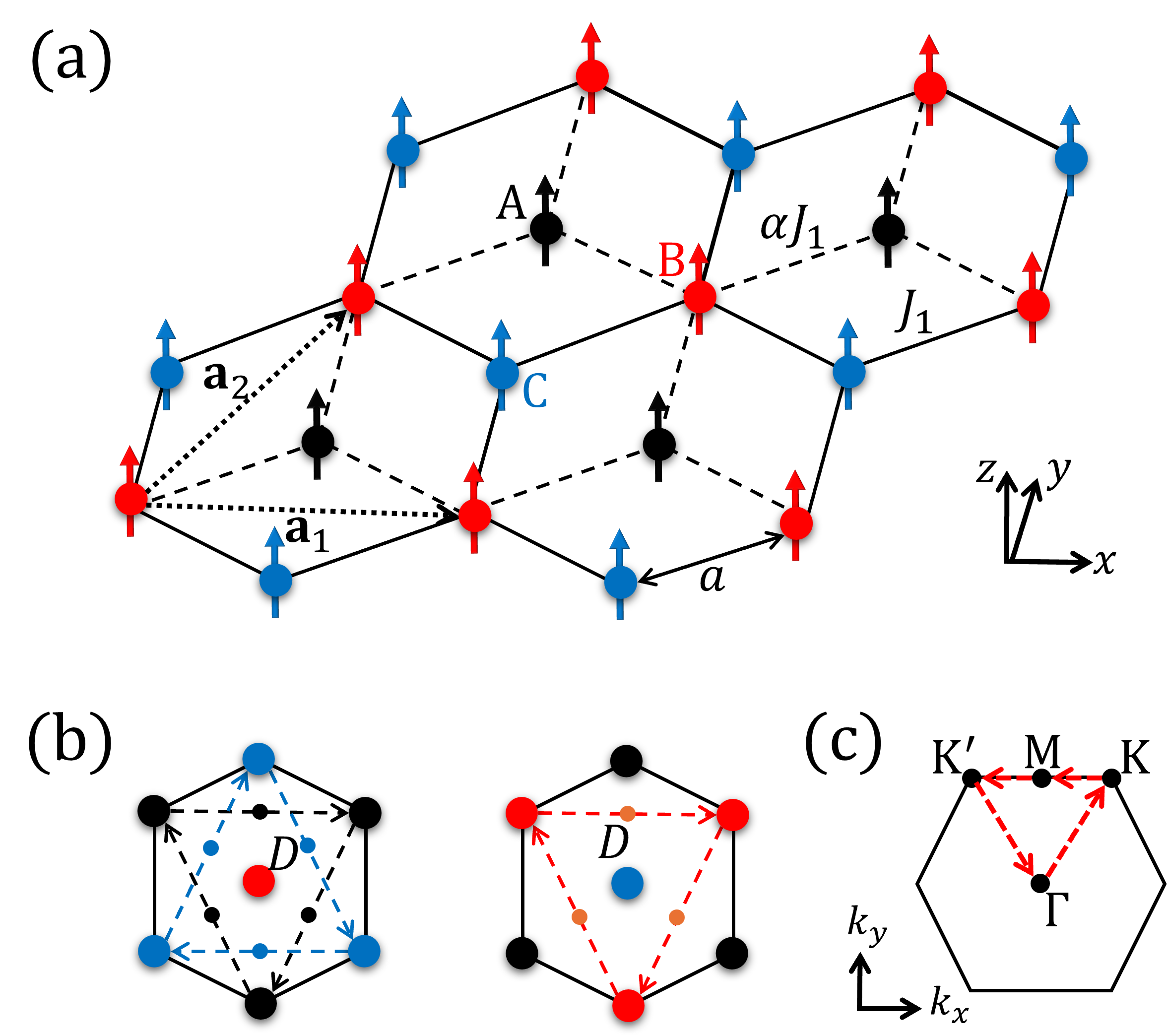}
\caption{(a) $\alpha$-T$_3$ lattice with the spins aligned along the $z$ axis. $\mybf{a}_1$ and $\mybf{a}_2$ are the lattice vectors. (b) DM interaction between next-nearest neighbors. (c) High symmetry points in the first Brillouin zone and path chosen for plotting the magnon band structure.}
\label{fig:fig1}
\end{figure}

We consider the $\alpha$-T$_3$ lattice shown in Figure \ref{fig:fig1}(a,b). The spin Hamiltonian is given by
\begin{align} \label{H1}
H =& -\sum_{\langle i,j \rangle} J_1^{ij} \, \mathbf{S}_i \cdot \mathbf{S}_j - J_2 \sum_{\langle\langle i,j \rangle\rangle}  \, \mathbf{S}_i \cdot \mathbf{S}_j  \nonumber \\
&+ \sum_{\langle\langle i,j \rangle\rangle} D \nu_{ij} \hat{z} \cdot (\mathbf{S}_i \times \mathbf{S}_j)- J_K \sum_{\langle i,j \rangle} (S_i^z)^2,
\end{align}
where $\langle .,. \rangle $ represents nearest neighbor sites, separated by a distance $a$, $\langle \langle .,. \rangle \rangle $ represents next-nearest neighbor sites, $\mathbf{S}_i$ represents the spin on site $i$, and $J_1^{ij}>0$ and $J_2>0$ represent the nearest and next-nearest neighbor coupling strengths, respectively. We set $J_1^{ij} = \alpha J_1$ for the coupling between sublattices A and B, and $J_1^{ij} = J_1$ for the coupling between sublattices B and C. The third term in Eq.\ (\ref{H1}) represents the out-of-plane DM interaction with $D \geq 0$ analogous to the honeycomb lattice \cite{PhysRevLett.127.217202}, $\nu_{ij} = +1$ for clockwise and $-1$ for counterclockwise hopping, and $\char`\^$ denoting a vector of unit length. $ J_K > 0 $ represents the anisotropy strength defining the easy axis. 
	
We utilize the linear spin-wave formalism \cite{Petit2011, Toth2015} to derive the tight-binding magnon Hamiltonian in momentum space. To this aim, we represent $H$ in terms of magnon operators by applying the lowest order Holstein-Primakoff transformation \cite{PhysRev.58.1098}
\begin{align} \label{E:HP_transformation}
\begin{aligned}
S_{i}^{+}&=S_{i}^x + iS_{i}^y\approx\sqrt{2S} b_{i},  \\
S_{i}^{-}&=S_{i}^x - iS_{i}^y\approx\sqrt{2S}b_{i}^{\dagger},\\
S_{i}^{z}&=(S-b_{i}^{\dagger}b_{i}),
\end{aligned}
\end{align}
where $b_{i}$ and $b_{i}^\dagger$ are the magnon annihilation and creation operators, respectively, and we assume $|\mybf{S}_i|=S$ for all $i$. Keeping only the quadratic terms in the magnon operators we then perform the Fourier transformation
\begin{equation} \label{E:magnon_FT}
b_{i}^{(\dagger)}=\frac{1}{\sqrt{L}}\sum_{\mybf{k}}\text{e}^{(-)i\mybf{t}_{i}\cdot \mybf{k}}b_{\mybf{k}}^{(\dagger)},
\end{equation}
where $b_{\mybf{k}}$ and $b_{\mybf{k}}^{\dagger}$ are the magnon annihilation and creation operators in Fourier space, respectively, $L$ is the number of unit cells, and $\mybf{t}_{i}$ is the position vector of site $i$. We obtain
\begin{equation} \label{E:H2}
H = \sum_{\mybf{k}} \Psi_{\mybf{k}}^\dagger h(\mybf{k}) \Psi_{\mybf{k}},
\end{equation}
where $\Psi_{\mybf{k}}$ is the magnon basis and $h(\mybf{k})=h_1(\mybf{k})+h_2(\mybf{k})+h_{D}(\mybf{k})+h_K$ with
\begin{align}
\begin{aligned}
h_1(\mathbf{k}) &=  J_1 S \begin{bmatrix}
				3\alpha  & -\alpha \Gamma_{1,\mathbf{k}} & 0 \\
				-\alpha \Gamma_{1,\mathbf{k}}^* & 3 (1+\alpha) & - \Gamma_{1,\mathbf{k}} \\
				0 & - \Gamma_{1,\mathbf{k}}^* & 3 
			\end{bmatrix}, \\
h_2(\mathbf{k}) &=  J_2 S (6-\Gamma_{2,\mathbf{k}})I_{3\times 3}, \\
h_D(\mathbf{k}) &=  D S \begin{bmatrix}
				\Gamma_{D,\mathbf{k}}  & 0 & 0 \\
				0 & \Gamma_{D,\mathbf{k}} & 0 \\
				0 & 0 & -\Gamma_{D,\mathbf{k}}
			\end{bmatrix}, \\
h_K &=  2 J_K S I_{3\times 3}
\end{aligned}
\end{align}
and the abbreviations
\begin{align}
\begin{aligned}
\Gamma_{1,\mathbf{k}} &= \exp{(iak_y)} + 2\cos{\left(\frac{\sqrt{3} a }{2}k_x\right)}\exp\left(-\frac{ia}{2}k_y\right), \\
\Gamma_{2,\mathbf{k}} &= 2\left(\cos(\sqrt{3} a k_x) + 2 \cos\left(\frac{\sqrt{3} a}{2} k_x\right) \cos\left(\frac{3a}{2} k_y\right)\right), \\
\Gamma_{D,\mathbf{k}} &= -2\left(\sin(\sqrt{3} a k_x) - 2 \sin\left(\frac{\sqrt{3} a}{2} k_x\right) \cos\left(\frac{3a}{2} k_y\right)\right).
\end{aligned}
\end{align}
$I_{3\times 3}$ is the three-dimensional identity matrix. We set $a=1$ nm in the following. 
	
Note that $H$ in Eq.\ (\ref{E:H2}) is restricted to quadratic order in the magnon operators, which is valid at low temperature. Higher-order corrections can be incorporated by including quadratic and higher terms in the magnon operators into the Holstein-Primakoff transformation and utilizing the methods proposed in Ref.\ \cite{PhysRevB.109.134415}. These corrections encompass thermally activated processes and spontaneous decay \cite{RevModPhys.85.219}. While the former are frozen out at low temperature, the latter broaden the magnon spectrum \cite{PhysRevB.101.024427}. Inclusion of higher-order corrections, such as magnon-magnon interactions, is out of the scope of this work. However, we anticipate that they do not alter the qualitative conclusions drawn from our analysis.

\subsection{Atomistic spin dynamics}
Spin dynamics simulations are conducted using the atomistic Landau-Lifshitz-Gilbert equation \cite{Landau:437299,1353448} 
\begin{align} \label{E:LLG}  
\frac{\partial \mybf{S}_i}{\partial t} = &-\frac{\gamma}{(1+\Delta^2)\mu_i} \mybf{S}_i \times \mybf{B}_i^{eff} \nonumber \\  
&-\frac{\gamma\Delta}{(1+\Delta^2)\mu_i} \mybf{S}_i \times (\mybf{S}_i \times \mybf{B}_i^{eff}),  
\end{align}  
where $\gamma$ is the gyromagnetic ratio, $\Delta$ is the Gilbert damping, $\mu_i$ is the spin moment, and $\mybf{B}_i^{eff}$ is the effective magnetic field (which depends on $H$ and at $T=0$ is given by $\mybf{B}_i^{eff} = \partial H / \partial \mybf{S}_i$). The first term in Eq.\ (\ref{E:LLG}) represents the quantum mechanical precession of spins, while the second term represents the damping that aligns the spin direction with an applied magnetic field. To numerically evolve a spin according to Eq.\ (\ref{E:LLG}), we use the Spirit code that adopts a fourth-order Runge-Kutta solver \cite{PhysRevB.99.224414}.

\subsection{Chern Number and Magnon Thermal Conductivity}
Topological phases of matter are well investigated for electrons \cite{Moore2010,RevModPhys.82.3045}. One example is the Chern insulator, which differs from a trivial insulator in that the Bloch states acquire along a closed path a non-zero Berry phase \cite{doi:10.1098/rspa.1984.0023} (which defines a non-zero Berry curvature). A non-trivial topology of Bloch states thus is characterized by a non-zero Chern number. This concept of topology is not exclusive to electrons but can also be applied to magnon Bloch states \cite{Owerre2016,10.1063/5.0041781,R2,R1}. In our case, the topology is driven by the DM interaction, as magnons acquire a phase from one site to another when $D \neq 0$, giving rise to the Berry curvature
\begin{equation} 
	\Omega_{n,\mathbf{k}}^z = -2 \hbar^2 \text{Im} \sum_{\substack{m=1, \\ m \ne n}}^3 \frac{v_x^{nm}(\mathbf{k}) v_y^{mn}(\mathbf{k})}{(\epsilon_{n,\mathbf{k}} - \epsilon_{m,\mathbf{k}})^2} \label{E:Omega}
\end{equation}
of the $n$-th magnon Bloch state. Here, $\hbar$ is Planck's constant, $v_i^{nm}(\mybf{k})=\langle n,\mybf{k} |\partial h(\mybf{k})/\partial k_i| m ,\mybf{k} \rangle$ are the matrix elements of the velocity operator in direction $i$, and $| n,\mybf{k} \rangle$ and $\epsilon_{n,\mybf{k}}$ are the $n$-th eigenstate and eigenvalue of $h(\mybf{k})$, respectively. The Chern number of the $n$-th magnon band is given by integration of $\Omega_{n,\mybf{k}}^z$ over the first Brillouin zone,
\begin{equation} 
	C_{n} = \frac{1}{2\pi} \int d^2 \mybf{k}\ \Omega_{n,\mybf{k}}^z. \label{E:chern}
\end{equation}
	
$C_n\ne 0$ indicates that the magnon thermal Hall conductivity of the $n$-th magnon band \cite{PhysRevB.109.134415} 
\begin{equation} \label{E:k_ij2}
{\kappa}_{xy}^n = - \frac{k_B^2T}{\hbar} \int \frac{d^2\mybf{k}}{4\pi^2} \Omega_{n,\mybf{k}}^z c_2(\epsilon_{n,\mybf{k}})
\end{equation}
is non-zero, where the temperature gradient is applied in the $y$ direction and the thermal current is measured in the $x$ direction. We denote the Boltzmann constant as $k_B$ and
\begin{equation} \label{E:c2}
c_2(\epsilon_{n,\mybf{k}}) = - \int^{\infty}_{\epsilon_{n,\mybf{k}}} d\epsilon\ 
\left(\frac{\epsilon}{k_BT}\right)^2 f_{B}'(\epsilon),
\end{equation}
where $f_B'(\epsilon)$ is the derivative of the Bose distribution function. The magnon thermal Hall conductivity then is given by $\kappa_{xy} = \sum_{n=1}^3 \kappa_{xy}^n$. The longitudinal thermal conductivity analogously is given by $\kappa_{xx}=\sum_{n=1}^3 \kappa_{xx}^n$ with
\begin{equation} \label{E:kappa_xx}
{\kappa}_{xx}^n = -\frac{\hbar}{T\eta} \sum_{m=1}^3\int \frac{d^2\mybf{k}}{4\pi^2} |v_x^{mm}(\mybf{k})|^2 \epsilon_{n,\mybf{k}}^2 f'_B (\epsilon_{n,\mybf{k}}),
\end{equation}
where $\eta$ is the magnon broadening. The magnon density of states is calculated through the retarded Green's function as
\begin{equation} \label{E:DOS}
\text{DOS($\epsilon_{n,\mybf{k}}$)} = -\frac{1}{\pi} \text{Im} \int \frac{d^2\mybf{k}}{4\pi^2} \text{Tr}\left[\frac{1}{\epsilon_{n,\mybf{k}} - h(\mybf{k}) +i\eta }\right].
\end{equation}
	
\begin{figure}
\includegraphics[width=\linewidth]{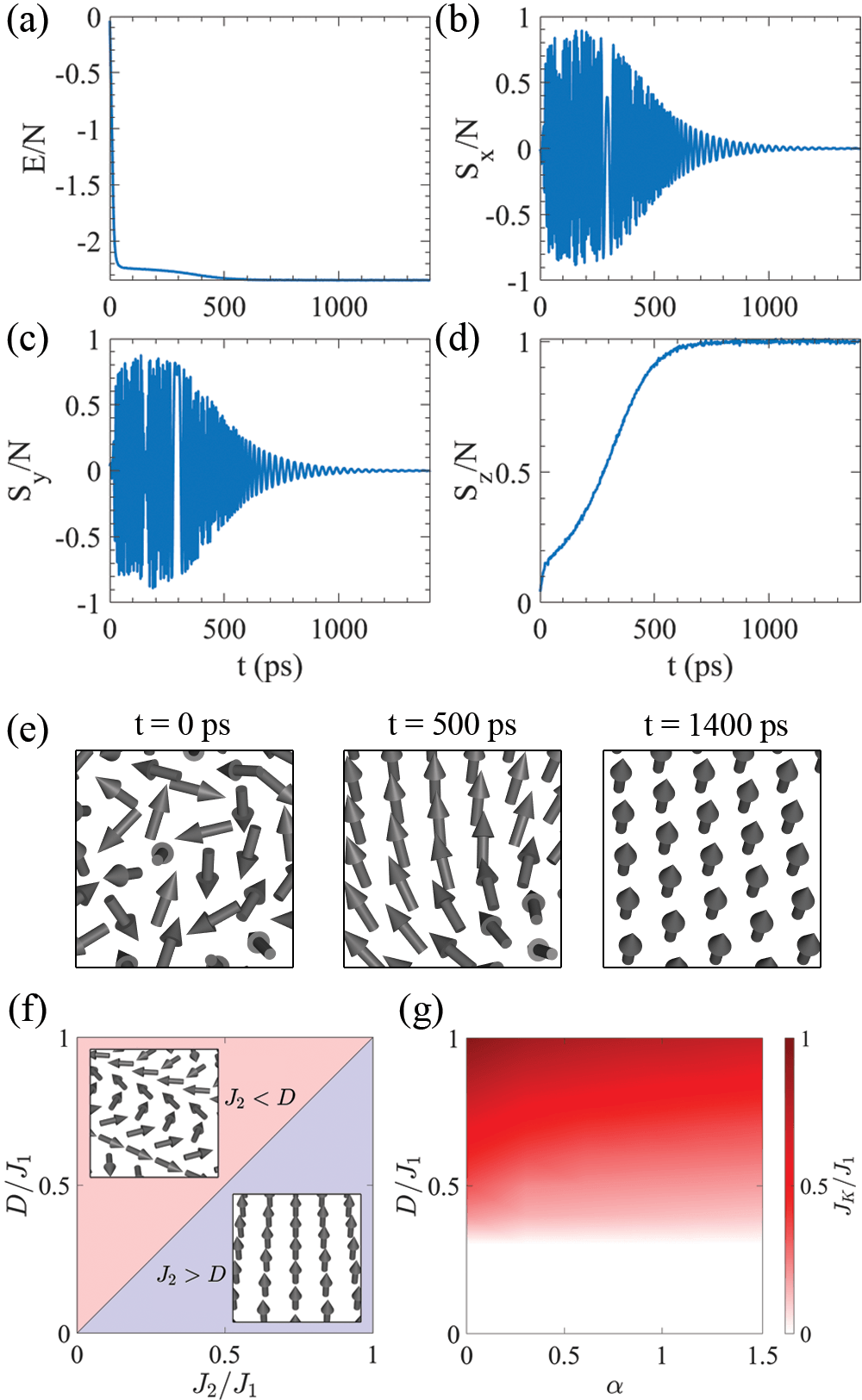}
\caption{(a) Energy density and (b-d) densities of the $x$, $y$, and $z$ spin components for $J_2/J_1 = D/J_1 = J_K/J_1 = 0$ and $\alpha=1.25$. $N$ is the number of lattice sites. (e) Snapshots of the spin configuration at times $0$, $500$, and $1400$ ps. (f) Magnetic phase diagram for $J_K/J_1=0$ and $0\leq\alpha\leq1.5$, showing ferromagnetic order for $J_2>D$ and spiral order for $J_2<D$. (g) Critical value of $J_K/J_1$ above which the system evolves to ferromagnetic order.}
\label{fig:fig2}
\end{figure}
	
\begin{figure}[b]
\includegraphics[width=1\linewidth]{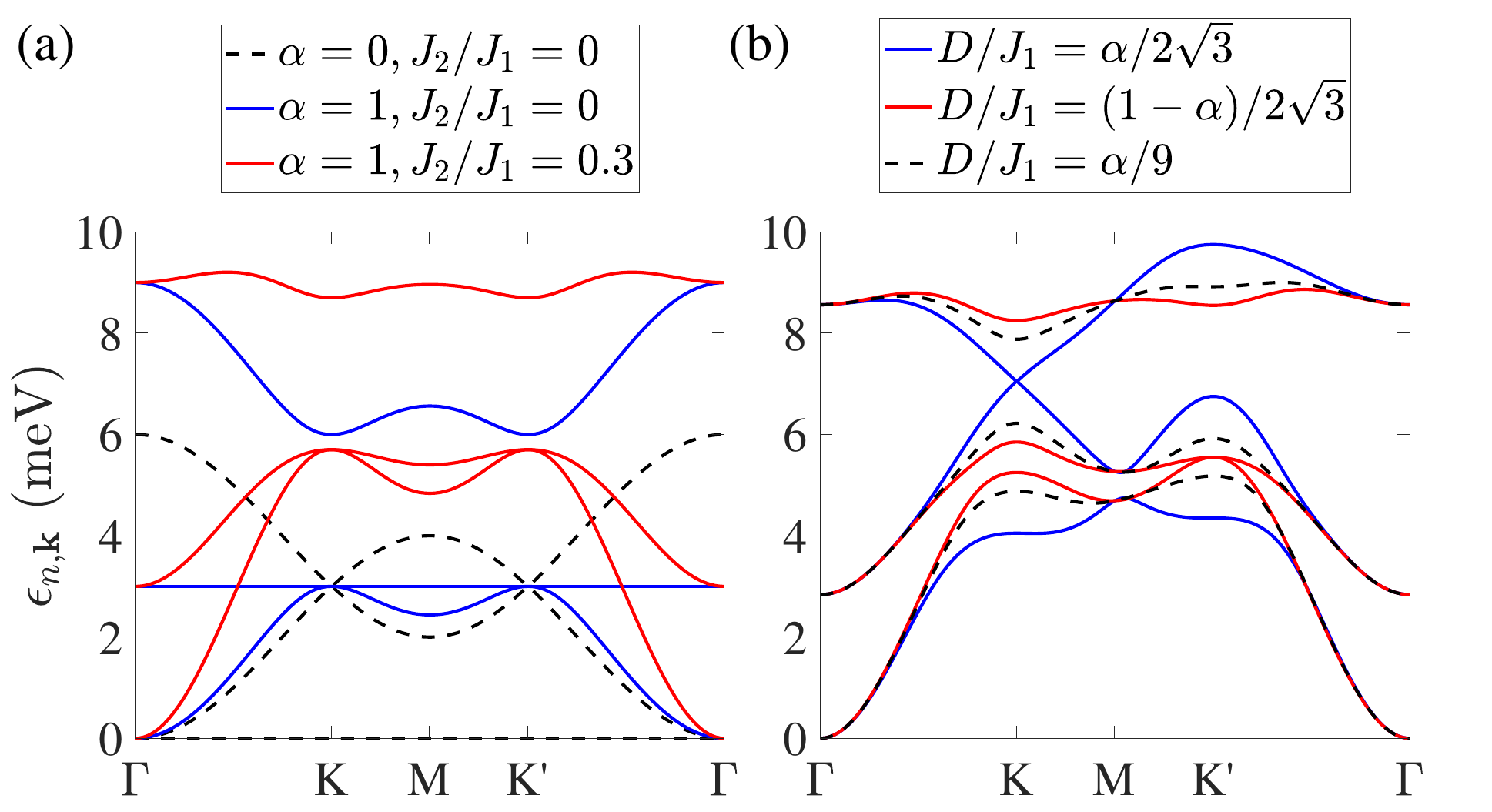}
\caption{(a) Magnon band structures at $D/J_1=0$ for the honeycomb ferromagnet at $J_2/J_1=0$ (black dashed lines) as well as the dice ferromagnet at $J_2/J_1=0$ (blue solid lines) and $J_2/J_1 = 0.3$ (red solid lines). (b) Magnon band structures at $J_2/J_1=0.3$ and $\alpha=0.9$ for different values of $D/J_1$. The high symmetry points are defined in Figure \ref{fig:fig1}(c).}
\label{fig:fig3}
\end{figure}
	
\section{Results and Discussion} \label{sec:results}
\subsection{Magnetic Order}
We conduct atomistic spin dynamics simulations using the Spirit code \cite{PhysRevB.99.224414} to determine the values of the parameters $\alpha$, $J_1$, $J_2$, $D$, and $J_K$ that support ferromagnetic order. They are performed for $10 \times 10 \times 1$ unit cells with periodic boundary conditions in the $x$ and $y$ directions. We solve Eq.\ (\ref{E:LLG}) using a Depondt solver, initiating the simulations with a random spin configuration at $t = 0$. The spins are allowed to evolve under a Gilbert damping of $0.05$ with a time step of $1$ fs. First, we examine the case $J_2/J_1 = D/J_1 = J_K/J_1 = 0$ to establish the effect of varying $\alpha$ from $0$ to $1.5$, finding for all values ferromagnetic order after a few ns. Figure \ref{fig:fig2}(a-e) illustrates the stability of the ferromagnetic order for $\alpha = 1.25$, as the energy density (energy per number of lattice sites) converges, the densities of the $x$ and $y$ spin components (spin per number of lattice sites) approach zero, and the density of the $z$ spin component approaches unity. Second, for $D/J_1 = J_K/J_1 = 0$ we vary both $\alpha$ and $J_2/J_1$ from $0$ to $1.5$, finding always ferromagnetic order. Finally, by fixing $J_2/J_1$ and varying $D/J_1$ from $0$ to $1$ we determine the critical value of $J_K/J_1$ required to realize ferromagnetic order for different $\alpha$. We obtain for $J_K/J_1=0$ spiral order as long as $D/J_1 > J_2/J_1$, as shown in Figure \ref{fig:fig2}(f). Figure \ref{fig:fig2}(g) shows the critical value of $J_K/J_1$ at $J_2/J_1 = 0.3$ for different values of $D/J_1$ and $\alpha$. As it does not significantly impact our qualitative analysis, we fix $J_2/J_1 = 0.3$ unless stated otherwise. Moreover, based on the above results, we constrain $0 \leq D/J_1 \leq 0.3$ and set $J_K/J_1 = 0$ to ensure ferromagnetic order.
	
\subsection{Band Structure and Topological Phase Diagram}
For simplicity, we set in the following $S=1$. The magnon band structure of the honeycomb ferromagnet at $J_2/J_1=D/J_1=0$ is shown by black dashed lines in Figure \ref{fig:fig3}(a), reproducing the results of Ref.\ \cite{PhysRevLett.127.217202}. The dice ferromagnet exhibits two dispersive bands (lower and upper) and one flat band (middle) for $J_2/J_1=D/J_1=0$. The lower and middle bands touch each other at the K- and K'-points, while the upper band is separated by a band gap of at least $3J_1 S$ (as the coordination number of sublattice B is twice that of sublattices A and C). $J_2/J_1\ne0$ makes the middle band dispersive even though no band gap opens to the lower band. Variation of $D/J_1$ and $\alpha$ can be used to open a band gap at the K- or K'-point, or even everywhere in the reciprocal space. At $D/J_1=\alpha/2\sqrt{3}$ the middle and upper bands touch each other at the K-point, at $D/J_1=(1-\alpha)/2\sqrt{3}$ the lower and middle bands touch each other at the K'-point, and at $D/J_1=\alpha/9$ the bands do not touch each other, which is demonstrated in Figure \ref{fig:fig3}(b) for $\alpha=0.9$. Note that a band gap opens not only for $D/J_1=\alpha/9$ but for countless relations between $D/J_1$ and $\alpha$.
	
\begin{figure}[t]
\includegraphics[width=\linewidth]{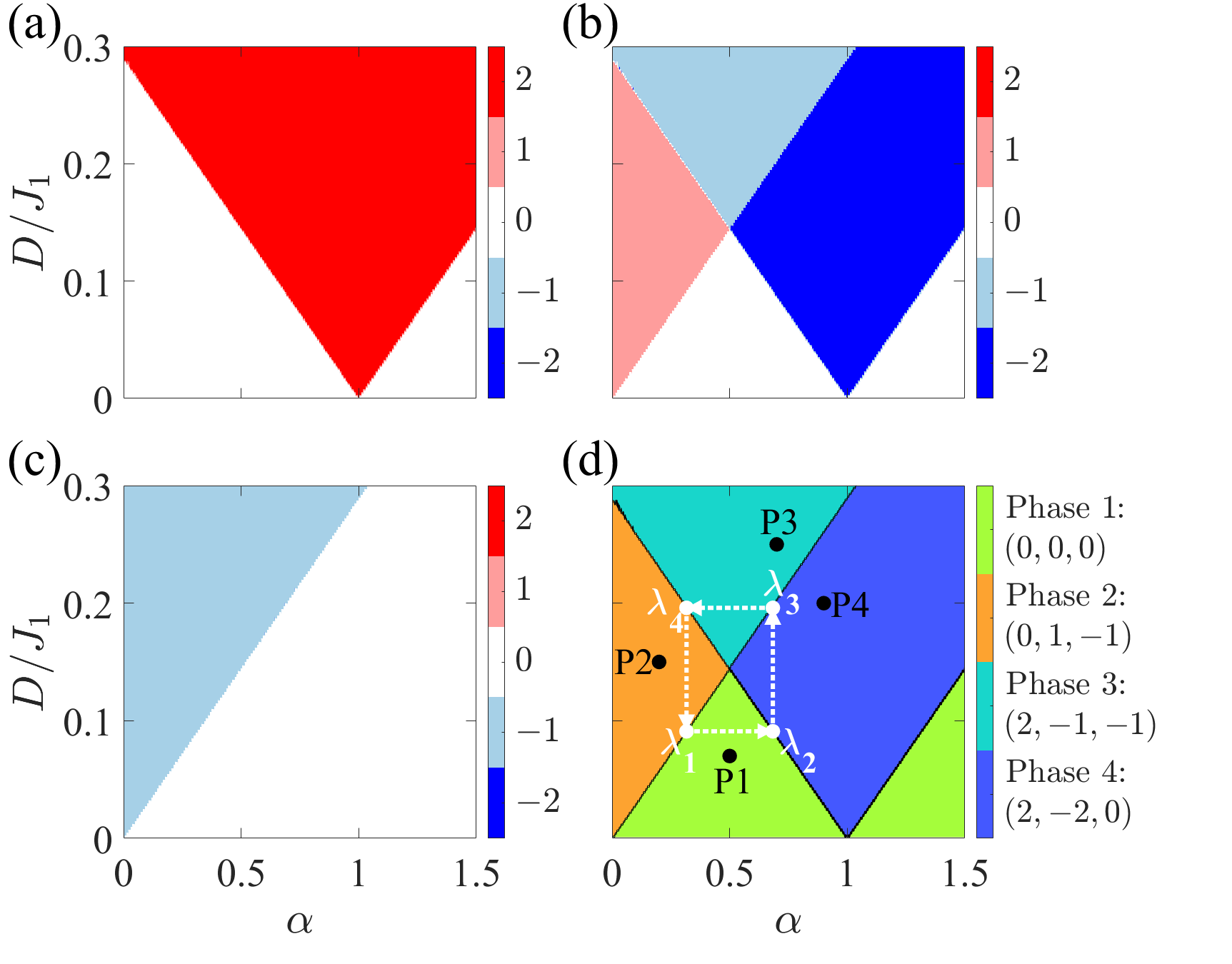}
\caption{Chern numbers of the (a) lower, (b) middle, and (c) upper magnon bands. (d) Phase diagram in the $\alpha\text{-}D/J_1$ plane. The points P1-P4 and the rectangular path with corners $\lambda_1$-$\lambda_4$ are chosen to analyze the thermal transport properties.}
\label{fig:fig4}
\end{figure}

Following Ref.\ \cite{PhysRevB.109.054412}, we define the topological phase $(C_1, C_2, C_3)$ through the Chern numbers $C_1$, $C_2$, and $C_3$ of the lower, middle, and upper magnon bands, see Figure \ref{fig:fig4}(a-c). The phase diagram in the $\alpha$-$D/J_1$ plane is shown in Figure \ref{fig:fig4}(d). We find one topologically trivial magnon insulator phase $(0,0,0)$ and three magnon Chern insulators phases $(0,1,-1)$, $(2,-1,-1)$, and $(2,-2,0)$. While $\alpha$ can open a band gap, the topologically trivial nature is maintained, indicating that the topological properties arise from the DM interaction. To analyze the behavior of $\kappa_{xy}$ for the four phases, we study the points $\text{P1}=(0.5,0.07)$, $\text{P2}=(0.2,0.15)$, $\text{P3}=(0.7,0.25)$, and $\text{P4}=(0.9,0.20)$ as well as the rectangular path with corners $\lambda_1=(0.32,0.09)$, $\lambda_2=(0.68,0.09)$, $\lambda_3=(0.68,0.20)$, and $\lambda_4=(0.32,0.20)$, see Figure \ref{fig:fig4}(d).
	
\begin{figure*}
\includegraphics[width=\linewidth]{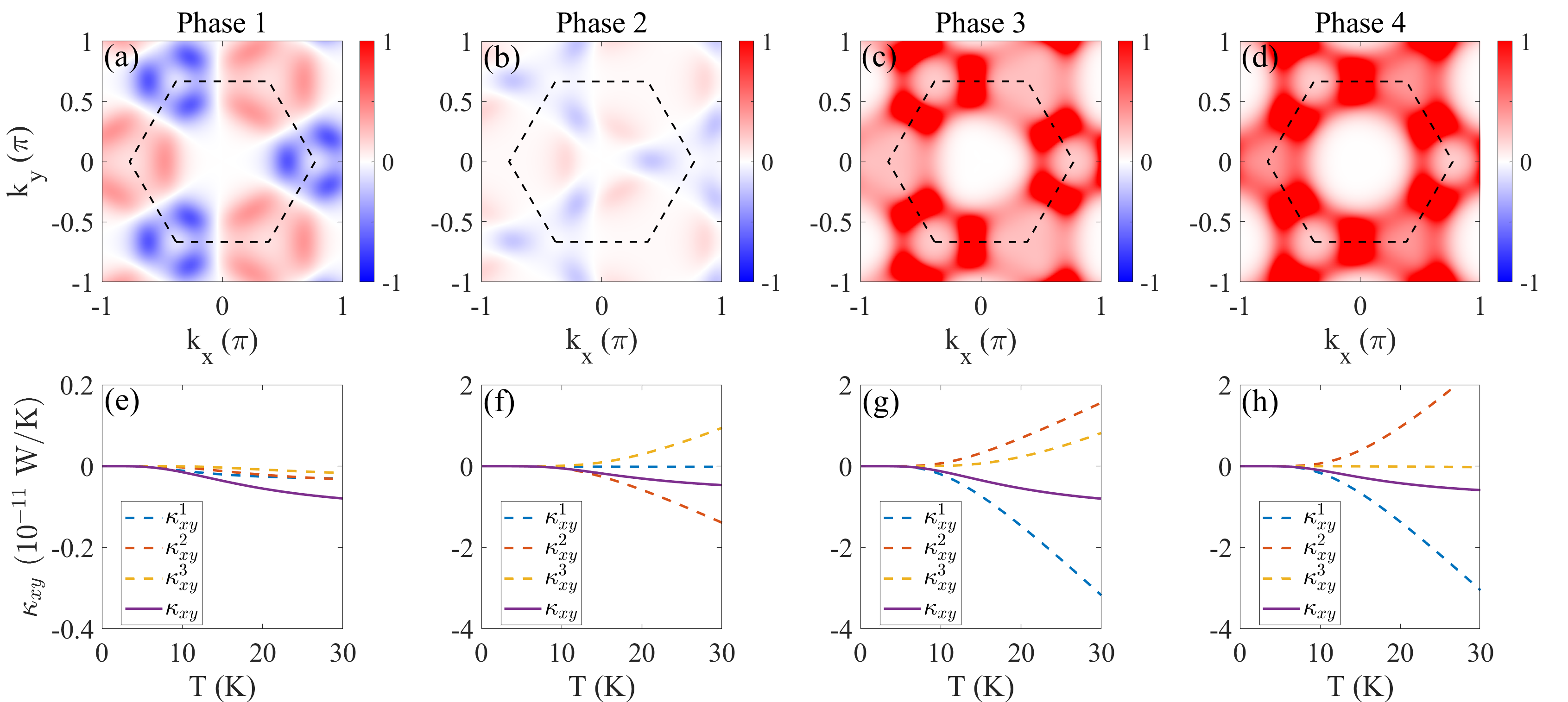}
\caption{(a-d) Berry curvature of the lowest magnon band and (e-h) magnon thermal Hall conductivity and band contributions as functions of the temperature. The black dashed hexagons in (a-d) represent the first Brillouin zone.}
\label{fig:fig5}
\end{figure*}
	
\begin{figure}
\includegraphics[width=\linewidth]{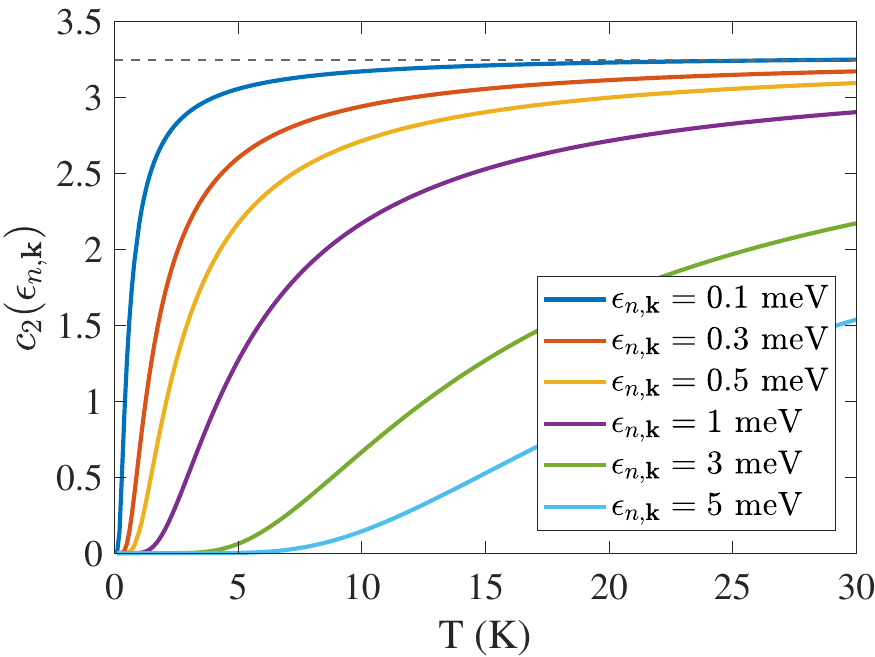}
\caption{$c_2(\epsilon_{n,\mybf{k}})$ as a function of the temperature at different $\epsilon_{n,\mybf{k}}$. The high temperature limit (dashed line) is $\pi^2/3$.}
\label{fig:fig6}
\end{figure}
	
\subsection{Thermal Transport Properties}
The electric Hall conductivity
\begin{equation}
-\frac{e^2}{\hbar}\sum_n\int \frac{d^2\mybf{k}}{4\pi^2}\Omega_{n,\mybf{k}}^z f_F(\epsilon_{n,\mybf{k}})
\end{equation}
of an electronic Chern insulator, where $f_F(\epsilon_{n,\mybf{k}})$ is the Fermi distribution function, is quantized and linked to the Berry curvatures of the electronic bands below the Fermi energy. As the Fermi energy can be adjusted by doping to lie within the band gap, the topological nature therefore can be probed by measuring the electric Hall conductivity. In contrast, magnons follow bosonic statistics and thus have no Fermi surface. Nevertheless, topological magnon phases can be explored by measuring $\kappa_{xy}$ \cite{doi:10.1126/science.1188260,doi:10.1126/sciadv.adk3539}.

\begin{figure}[ht!]
	\includegraphics[width=\linewidth]{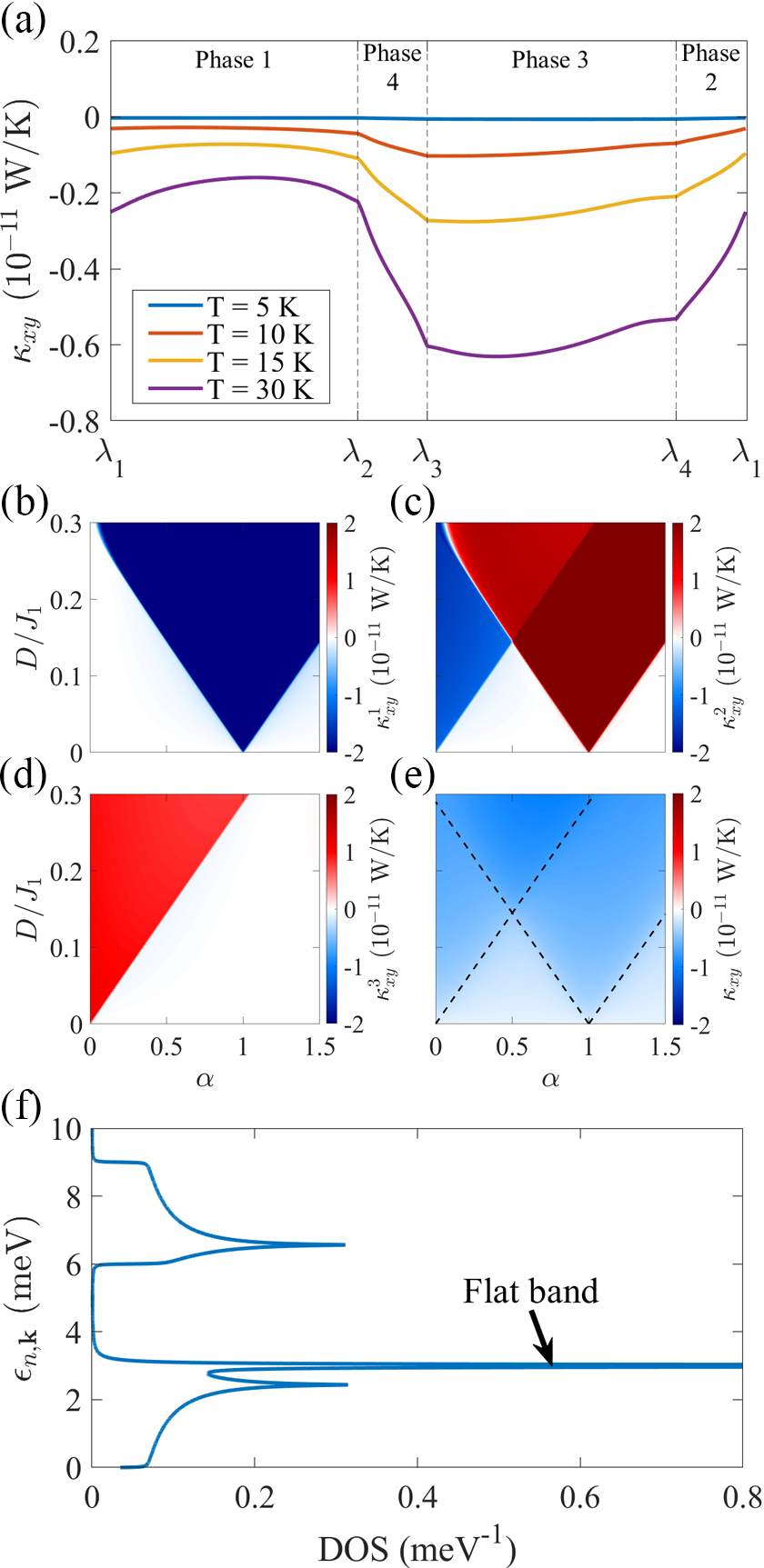}
	\caption{(a) Magnon thermal Hall conductivity along the rectangular path defined  in Figure \ref{fig:fig4}(d) at different temperatures. Magnon thermal Hall conductivities of the (b) lower, (c) middle, and (d) upper magnon bands as well as (e) total magnon thermal Hall conductivity in the $\alpha\text{-}D/J_1$ plane at $T = 30$ K. The black dashed lines in (e) mark the topological phase boundaries. (f) Magnon density of states for $J_2/J_1=D/J_1=0$. The divergence at $ \epsilon_{n,\mathbf{k}} = 3$ meV corresponds to the flat band in Figure \ref{fig:fig3}(a).}
	\label{fig:fig7}
\end{figure}

A necessary condition to obtain $\kappa_{xy}\ne 0$ is the absence of effective time-reversal symmetry, $\Omega_{n,\mybf{k}}^z = -\Omega_{n,-\mybf{k}}^z$ (combination of time-reversal symmetry and $\pi$ rotation of the spin) \cite{PhysRevB.99.014427}. In our case, we have neither effective time-reversal symmetry nor inversion symmetry due to $D\ne0$ and therefore can expect $\kappa_{xy}\ne 0$. The magnon thermal conductivity tensor for the magnetic point group $3m'1$ (compatible with ferromagnetism) has the form \cite{PhysRevB.92.155138}
	\begin{equation}
		\mybf{\kappa} = \begin{bmatrix}
			\kappa_{xx} & \kappa_{xy} & 0 \\ 
			-\kappa_{xy} & \kappa_{xx} & 0 \\ 
			0 & 0 & \kappa_{zz}
		\end{bmatrix}.
	\end{equation}
Note that magnon bands with $C_n=0$ can result in $\kappa_{xy}\ne 0$ as long as there is no effective time-reversal symmetry and $\Omega_{n,\mybf{k}}^z\ne 0$.

The Berry curvature of the lowest magnon band, $\Omega_{1,\mathbf{k}}^z$, is shown in Figure \ref{fig:fig5} for the four phases together with $\kappa_{xy}^n$ and $\kappa_{xy}$ as functions of $T$. It changes in phases 1 and 2 between negative and positive values across the reciprocal space, resulting in $C_1=0$, but is always positive in phases 3 and 4, resulting in $C_1>0$. Both $\kappa_{xy}^n$ and $\kappa_{xy}$ change monotonously for increasing $T$. Due to Eq.\ (\ref{E:k_ij2}), magnon bands with higher $|C_n|$ contribute more to $\kappa_{xy}$. $C_2=-C_3$ in phase 2 implies that $\kappa_{xy}^2$ and $\kappa_{xy}^3$ have opposite signs, see Figure \ref{fig:fig5}(f), and $C_1=2$ in phases 3 and 4 results in identical $\kappa_{xy}^1$, see Figure \ref{fig:fig5}(g, h). However, $\kappa_{xy}$ turns out to be negative in all the phases. Furthermore, magnon bands with the same $C_n$ can contribute differently to $\kappa_{xy}$. For instance, $C_2=C_3=-1$ in phase 3 but $|\kappa_{xy}^2|>|\kappa_{xy}^3|$, see Figure \ref{fig:fig5}(g). The reason is that Eq.\ (\ref{E:k_ij2}) connects $\kappa_{xy}^n$ to $c_2(\epsilon_{n,\mybf{k}})$, which decreases for increasing $\epsilon_{n,\mybf{k}}$ at given $T$, as shown in Figure \ref{fig:fig6}, thus acting similar to an activation function, that is, the contributions of higher magnon bands to $\kappa_{xy}$ increase as $T$ increases.
	
Figure \ref{fig:fig7}(a) shows $\kappa_{xy}$ at different $T$ along the rectangular path defined in Figure \ref{fig:fig4}(d). The path crosses all phases. $D/J_1$ is constant and $\alpha$ varies from $0.32$ to $0.68$ in path sections $\lambda_1\to\lambda_2$ and $\lambda_3\to\lambda_4$, while $\alpha$ is constant and $D/J_1$ varies from $0.009$ to $0.20$ in path sections $\lambda_2\to\lambda_3$ and $\lambda_4\to\lambda_1$. We observe that $|\kappa_{xy}|$ is generally lowest in phase 1 and highest in phase 3. It also is generally higher in phase 3 than in phase 4 (though $C_1=2$ in both phases) due to the fact that $C_2=-1$ in phase 3 results in less reduction than $C_2=-2$ in phase 4. While $\kappa_{xy}$ does not change sign at the topological phase boundaries, kinks are evident. They become more pronounced at higher $T$, as higher magnon bands become activated. Figure \ref{fig:fig7}(b-e) shows $\kappa_{xy}^n$ and $\kappa_{xy}$ in the $\alpha\text{-}D/J_1$ plane at $T = 30$ K. $\kappa_{xy}^n$ exhibits sharp changes at the topological phase boundaries in agreement with the changes of $C_n$, see Figure \ref{fig:fig4}(a-c). The topological phase boundaries are less sharp in the case of $\kappa_{xy}$ but still distinctively visible. At the flat band observed in Figure \ref{fig:fig3}(a) for $J_2/J_1=D/J_1=0$ the density of states diverges, as illustrated in Figure \ref{fig:fig7}(f) with $\eta=0.01J_1$. One may assume that the divergent density of states enhances the thermal transport, however, both $\kappa_{xx}^2$ and $\kappa_{xy}^2$ turn out to be zero, reflecting the vanishing group velocity. 
	
\section{Conclusion} \label{sec:summary}
We have explored the thermal transport properties of magnons on the $\alpha$-T$_3$ lattice. Atomistic spin dynamics simulations in the absence of DM interaction demonstrate always ferromagnetic order. The DM interaction promotes spiral order, while next-nearest neighbor hopping and easy-axis anisotropy promote ferromagnetic order. In the case of ferromagnetic order, we have identified one topologically trivial magnon insulator phase and three magnon Chern insulator phases. Magnon bands with higher $|C_n|$ contribute more to $\kappa_{xy}$ than those with lower $|C_n|$, but even magnon bands with $C_n=0$ give finite contributions. The topologically trivial magnon insulator phase exhibits the lowest $|\kappa_{xy}|$ and the magnon Chern insulator phase $(2,-1,-1)$ with the highest $|C_1|$ and lowest $|C_2|$ exhibits the highest $|\kappa_{xy}|$. While $\kappa_{xy}$ does not change sign at the topological phase boundaries, distinct changes are observed in the magnitude, which can be attributed to the magnon edge states present in the magnon Chern insulator phases.

In flat bands the quasiparticle interaction dominates over the kinetic energy, which may promote exotic many-body phenomena such as strongly correlated phases \cite{Paschen2021}, as observed for the Dice lattice \cite{PhysRevLett.108.045306}, and Bose-Einstein condensation in bosonic flat bands \cite{PhysRevB.82.184502}. Our results for non-interacting magnons (at low temperature) show that flat bands do not contribute to $\kappa_{xx}$ and $\kappa_{xy}$ due to vanishing group velocities. In contrast, at high temperatures magnons are inherently strongly interacting bosons. For this situation, our results provide a foundation for future research on how strong interaction in bosonic flat bands influences the topology and possible strongly correlated phases.

A ferromagnet with $\alpha$-T$_3$ lattice can be experimentally realized using a bilayer of the ferromagnetic honeycomb monolayers CrI$_3$ \cite{Huang2017}, VI$_3$ \cite{PhysRevMaterials.3.121401}, CrSiTe$_3$ \cite{V.Carteaux_1995}, CrGeTe$_3$ \cite{PhysRevB.95.245212} or Fe$_3$GeTe$_2$ \cite{Fei2018}, arranged in a way that a lattice site of one monolayer aligns with the center of the honeycomb of the other monolayer. Another possibility is a trilayer structure of cubic ferromagnets, such as iron oxides \cite{n1}, grown along the (111) direction. Nanomagnets \cite{PhysRevLett.106.057209,Canals2016,Leo2018,n2} also offer a promising avenue, as the structure can be flexibly designed.

\begin{acknowledgments}
The authors thanks A.\ S.\ Alzahrani for helpful discussions. This work was supported by King Abdullah University of Science and Technology.
\end{acknowledgments}

\vfill
	
\bibliography{bib}	
\end{document}